\documentclass[journal]{IEEEtran}
\usepackage[utf8]{inputenc}
\ifCLASSINFOpdf
  \usepackage[pdftex]{graphicx}
  \usepackage{epstopdf}
  \graphicspath{{Figures/}}
\else
\fi
\usepackage{amsmath}
\DeclareMathOperator*{\argmin}{\arg\!\min}

\newcommand{\norm}[1]{\left\lVert#1\right\rVert}
\usepackage{listings}
\usepackage{gensymb}

\begin{document}
%
% paper title
% Titles are generally capitalized except for words such as a, an, and, as,
% at, but, by, for, in, nor, of, on, or, the, to and up, which are usually
% not capitalized unless they are the first or last word of the title.
% Linebreaks \\ can be used within to get better formatting as desired.
% Do not put math or special symbols in the title.
\title{Regression-based Intra-prediction for Image and Video Coding}
%
%
% author names and IEEE memberships
% note positions of commas and nonbreaking spaces ( ~ ) LaTeX will not break
% a structure at a ~ so this keeps an author's name from being broken across
% two lines.
% use \thanks{} to gain access to the first footnote area
% a separate \thanks must be used for each paragraph as LaTeX2e's \thanks
% was not built to handle multiple paragraphs
%

\author{Carlo~Noel~Ochotorena,~\IEEEmembership{Member,~IEEE,}~and~Yukihiko~Yamashita,~\IEEEmembership{Member,~IEEE}% <-this % stops a space
	\thanks{C. N. Ochotorena is a student of the Graduate School of Science and Engineering, Tokyo Institute of Technology, Tokyo, Japan and a faculty member of De La Salle University, Manila, Philippines (e-mail: carlo.ochotorena@dlsu.edu.ph)}% <-this % stops a space
	\thanks{Y. Yamashita is with the Graduate School of Science and Engineering, Tokyo Institute of Technology, Tokyo, Japan (e-mail: yamasita@ide.titech.ac.jp)}% <-this % stops a space
	\thanks{This paper has supplementary downloadable material available at http://ieeexplore.ieee.org/, provided by the author. The material includes the source codes to generate the figures and additional data. The total size of the material is 17 MB including all the necessary images. Contact carlo.ochotorena@dlsu.edu.ph for further questions about this work.}
	\thanks{Manuscript received XXXXXXXXXXXX, 2016.}}

\maketitle

% As a general rule, do not put math, special symbols or citations
% in the abstract or keywords.
\begin{abstract}
By utilizing previously known areas in an image, intra-prediction techniques can find a good estimate of the current block. This allows the encoder to store only the error between the original block and the generated estimate, thus leading to an improvement in coding efficiency. Standards such as AVC and HEVC describe expert-designed prediction modes operating in certain angular orientations alongside separate DC and planar prediction modes. Being designed predictors, while these techniques have been demonstrated to perform well in image and video coding applications, they do not necessarily fully utilize natural image structures. In this paper, we describe a novel system for developing predictors derived from natural image blocks. The proposed algorithm is seeded with designed predictors (e.g. HEVC-style prediction) and allowed to iteratively refine these predictors through regularized regression. The resulting prediction models show significant improvements in estimation quality over their designed counterparts across all conditions while maintaining reasonable computational complexity. We also demonstrate how the proposed algorithm handles the worst-case scenario of intra-prediction with no error reporting.
\end{abstract}

% Note that keywords are not normally used for peerreview papers.

% For peer review papers, you can put extra information on the cover
% page as needed:
% \ifCLASSOPTIONpeerreview
% \begin{center} \bfseries EDICS Category: 3-BBND \end{center}
% \fi
%
% For peerreview papers, this IEEEtran command inserts a page break and
% creates the second title. It will be ignored for other modes.
\IEEEpeerreviewmaketitle

\section{Introduction}
% The very first letter is a 2 line initial drop letter followed
% by the rest of the first word in caps.
% 
% form to use if the first word consists of a single letter:
% \IEEEPARstart{A}{demo} file is ....
% 
% form to use if you need the single drop letter followed by
% normal text (unknown if ever used by the IEEE):
% \IEEEPARstart{A}{}demo file is ....
% 
% Some journals put the first two words in caps:
% \IEEEPARstart{T}{his demo} file is ....
% 
% Here we have the typical use of a "T" for an initial drop letter
% and "HIS" in caps to complete the first word.
\IEEEPARstart{I}{mages} and videos represent vast amounts of digital information. If left uncompressed, this information will correspond to a large storage footprint. Fortunately, a lot of the information in images is redundant. This allows for the use of transforms such as the DCT \cite{Ahm1974} or DWT \cite{Ant1992} to map image blocks into a more compact representation. This process allows a coder to make use of the correlation between neighboring pixels within a block to compress the data. However, this process does not take into consideration neighboring blocks. Intra-prediction schemes make use of these adjacent blocks to create an estimate of the current block to further compact the representation. In particular, previously encoded blocks can be used to form a good estimate of the current block.

A key problem in intra-prediction is finding a good way to estimate pixels from already known pixels. The simplest form of intra-prediction, used in the JPEG standard, is to simply assert that the mean of the current patch should be close to the mean of the previous patch \cite{Wal1992}. This form of intra-prediction amounts to no bit cost as only one prediction mode is used exclusively. On the other hand, the prediction, itself, may be quite inaccurate. More recent coding standards such as H.264/AVC \cite{Sul2005} and H.265/HEVC \cite{Lai2012} make use of more than one prediction mode to estimate the current patch. In such a case, the choice of prediction modes has to be encoded explicitly. There exists a clear trade-off between the number of prediction modes and the quality of the estimate.

Many studies have tackled the task of improving intra-prediction. On one hand, work is being done in efficiently guessing the next prediction mode to be used \cite{Wan2007,Zha2011a,Lar2010}. If these estimates are correct, then the number of bits needed to describe the next prediction mode can be reduced. Another area of interest is improving the prediction process itself. Many researchers have investigated this particular aspect of intra-prediction leading to a wide array of solutions \cite{An2008,Bal2007,Dai2008,Kam2013,Kon2004,Pia2010,Sun2012,Tan2013}.

In the AVC and HEVC standards, the intra-prediction modes are hand-designed to exploit natural image structures. Often, this is a combination of a zeroth-order DC prediction, a first-order planar prediction, and several angular prediction modes \cite{Lai2012}. In contrast, some effort has been made towards learning the predictors, instead \cite{Gar2010,Zha2011}. Based on these studies, there is a general improvement that can be obtained through the learning process. We hypothesize that the conventional designed predictors cannot fully utilize image structures as compared to these linear models.

\begin{figure}[t!]
	\centering
	\includegraphics[width=0.8\linewidth]{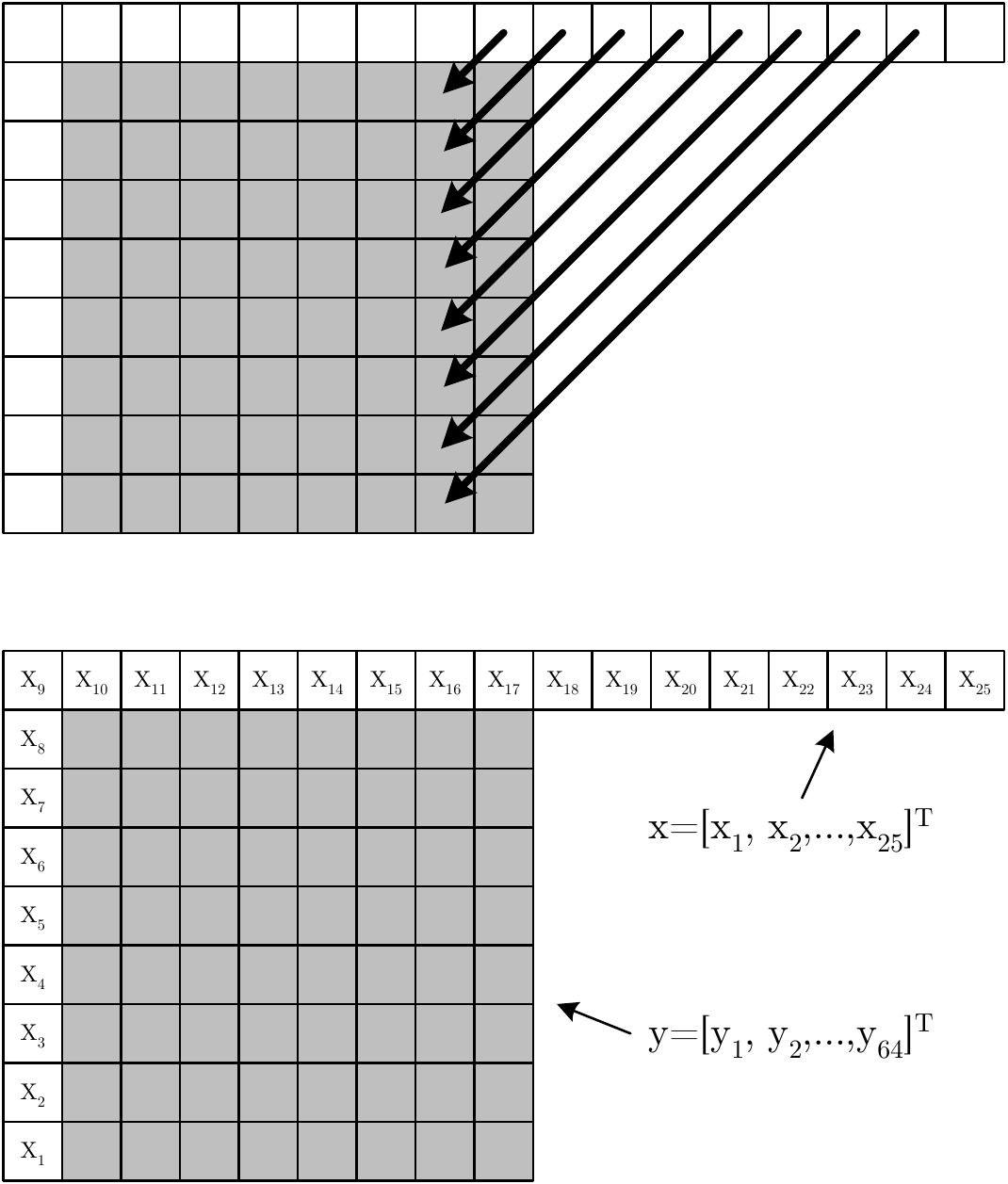}
	\caption{Intra-prediction can be seen as a pixel replication process (top) or treated as a mapping problem between two vectors (bottom).}
\end{figure}

In this work, we present a novel approach to intra-prediction. Our work is similar to \cite{Gar2010} and \cite{Zha2011} in that it utilizes training patches to train a linear mapping between the known pixels and the pixels to be predicted. However, unlike these approaches, we perform an iterative refinement process similar to that of the k-means clustering approach. We determine which predictor provides the best estimation of each patch and cluster these patches based on the best predictors. A new linear mapping is then calculated for each cluster. As the groupings may shift over time, this process is repeated iteratively. Once this offline learning process is completed, prediction for a single block can be carried out using a single matrix multiplication step.

As we demonstrate in this paper, the proposed algorithm presents several advantages over conventional prediction schemes when applied to images. Under the best conditions, we observe a significant improvement in estimates compared to conventional techniques. More interestingly, we present the results for the worst-case condition where only the top-left block is encoded explicitly and all other blocks are generated through intra-prediction with no error reporting. Through this, we illustrate a key feature of our approach. The results we report in this paper focus on images for the purpose of isolating the intra-prediction performance. However, our approach is easily extensible towards video coding applications as well. 

\section{Regression-based Intra-prediction (RIP)}

Intra-prediction algorithms depend on previously encoded portions of an image to form an estimate. While this process could theoretically involve all previously decoded pixels and blocks, it is often restricted to the bounding neighbors to reduce complexity (see Fig. 1). In the case of HEVC angular modes, every pixel in the current block is estimated as a weighted average of two pixels \cite{Lai2012}. Planar prediction increases this to a four point interpolation problem. DC prediction performs a simple average of all bounding pixels. In any of these cases, it is clearly defined that the set of bounding pixels are used to generate the block estimate.

From a mathematical perspective, this can be seen as a simple mapping task
\begin{equation}
\hat{y} = Mx
\end{equation}
where $x$ describes a column vector of all bounding pixels and $\hat{y}$ describes a column vector of all pixel predictions. In the $8 \times 8$ case (as seen in Fig. 1), this results in a 25-dimensional input vector mapping to a 64-dimensional output vector. In designed intra-prediction systems, such as the one used by HEVC, the prediction map $M$ appears to be a highly sparse matrix allowing for very efficient computation.

Given the above formulation, the intra-prediction problem can then be expressed as
\begin{equation}
\tilde{p} = \argmin_{p}{\norm{y-M_px}_2}
\end{equation}
where $y$ describes the true values of the block pixels and $M_p$ is used to describe the $p$-th prediction mode. The result $\tilde{p}$ denotes the prediction mode where minimum error is obtained.

Designed intra-prediction systems can then make use of the mode information during the encoding process. However, as we have hypothesized previously, these predictors may not fully exploit image structures. To address this, we group together similar blocks and, instead, attempt to learn the predictors for each group. A trivial manner of grouping these blocks is to make use of the associated predictors $\tilde{p}$ for each patch. It is natural to assume that if a group of patches use the same prediction mode, they are likely to behave similarly. Formally, we define matrices $X_j$ and $Y_j$ for the $j$-th group:
\begin{gather}
X_j = \lbrace x_i \mid \argmin_{p}{\norm{y_i-M_px_i}_2 = j}\rbrace\\
Y_j = \lbrace y_i \mid \argmin_{p}{\norm{y_i-M_px_i}_2 = j}\rbrace
\end{gather}

At this point, we utilize the target patches $y$ and not the estimates $\hat{y}$. A new predictor can easily by found using least-squares regression:
\begin{equation}
M_j = \argmin_{M}{\sum_{\substack{x_i\in X_j\\y_i\in Y_j}}\norm{y_i-Mx_i}_2}\\
\end{equation}

\begin{equation}
M_j = Y_jX_j^T(X_jX_j^T)^{-1}
\end{equation}

Experimentally, we have found that more stable mappings can be obtained when using an additional Tikhonov regularization term:
\begin{equation}
M_j = Y_jX_j^T(X_jX_j^T + \lambda I)^{-1}
\end{equation}

After all mappings have been learned, the resulting set of predictors should perform better than the original set. However, the patch groupings may also change with this new set of predictors. Intuitively, we can repeat the grouping and training process much like the k-means clustering approach. This allows us to construct progressively improving groups and their corresponding predictors. For all the experiments in this paper, a fixed number of 100 iterations is used. A description of the complete algorithm can be seen in Table 1.

\begin{table}[t!]
\centering
\caption{Pseudocode of the Proposed Algorithm}
\begin{tabular}{l}
\hline
initialize $M_j$ with the $j$-th conventional intra-prediction mode\\
until convergence\\
\hspace{0.5cm}find $X_j$ and $Y_j$ using (3) and (4)\\
\hspace{0.5cm}update $M_j$ using (7)\\
end\\
\hline
\end{tabular}
\end{table}

\section{Experiments}
In this work, we carry out two sets of experiments on our proposed Regression-based Intra-prediction (RIP). In the first series of tests, we observe the effect of increasing the number of angular modes. Our second test aims at comparing the RIP mappings with the HEVC prediction modes for varying block sizes. It should be emphasized at this point that our goal is to characterize the performance of the predictors themselves so coding-specific techniques such as quadtrees are excluded. Likewise, reference sample smoothing and boundary smoothing are not utilized as our primary metric is the error between the original and estimated blocks and not the perceptual quality of the resulting image.

\subsection{Effect of the Number Angular Modes}
While the effect of the number of angular modes has been characterized for designed intra-prediction schemes, this may not necessary apply in RIP. To verify this, we manually create a set of angular intra-prediction modes using the two-pixel interpolation scheme in HEVC. For simplicity, we divide the span between 45\degree\,and 225\degree\,into a set number of uniform angles. Specifically, we test the performance with 5, 9, 13, 17, 21, 25, 29, and 33 angular modes, chosen specifically to preserve the horizontal, vertical, and 45\degree\,diagonal orientations.

For the purposes of training, random patches are collected from each image in the \emph{Kodak PhotoCD PCD0992} image set. The RIP models are then obtained by applying the proposed refinement algorithm to the original intra-predictors using the training patches to facilitate the regression process. Images outside the training set are used to evaluate the performance of the original and RIP predictors. In particular, we use the well-known \emph{Lena}, \emph{Peppers}, and \emph{Mandrill} images to arrive at a quantitative analysis of the techniques. Without loss in generality, we perform prediction only on the luminance channel of the image.

To evaluate the different predictor sets, we prepared two sets of tests operating on an $8 \times 8$ block size. A best-case scenario test is performed where all inputs to the patch prediction algorithm are derived from the original data. In this manner, we obtain the best possible prediction assuming lossless encoding of all previous patches. On the opposite end of the spectrum, we test the behavior in the worst possible scenario. If we assume that the encoder has no bits to allocate for the prediction errors, then the image is reconstructed solely using the predictions from the previous patches as well as the information from the top-left patch. While it may be counter-intuitive to include such a test, we would like to demonstrate a unique characteristic of RIP through this.

\begin{figure}[t!]
	\centering
	\includegraphics[width=0.333\linewidth]{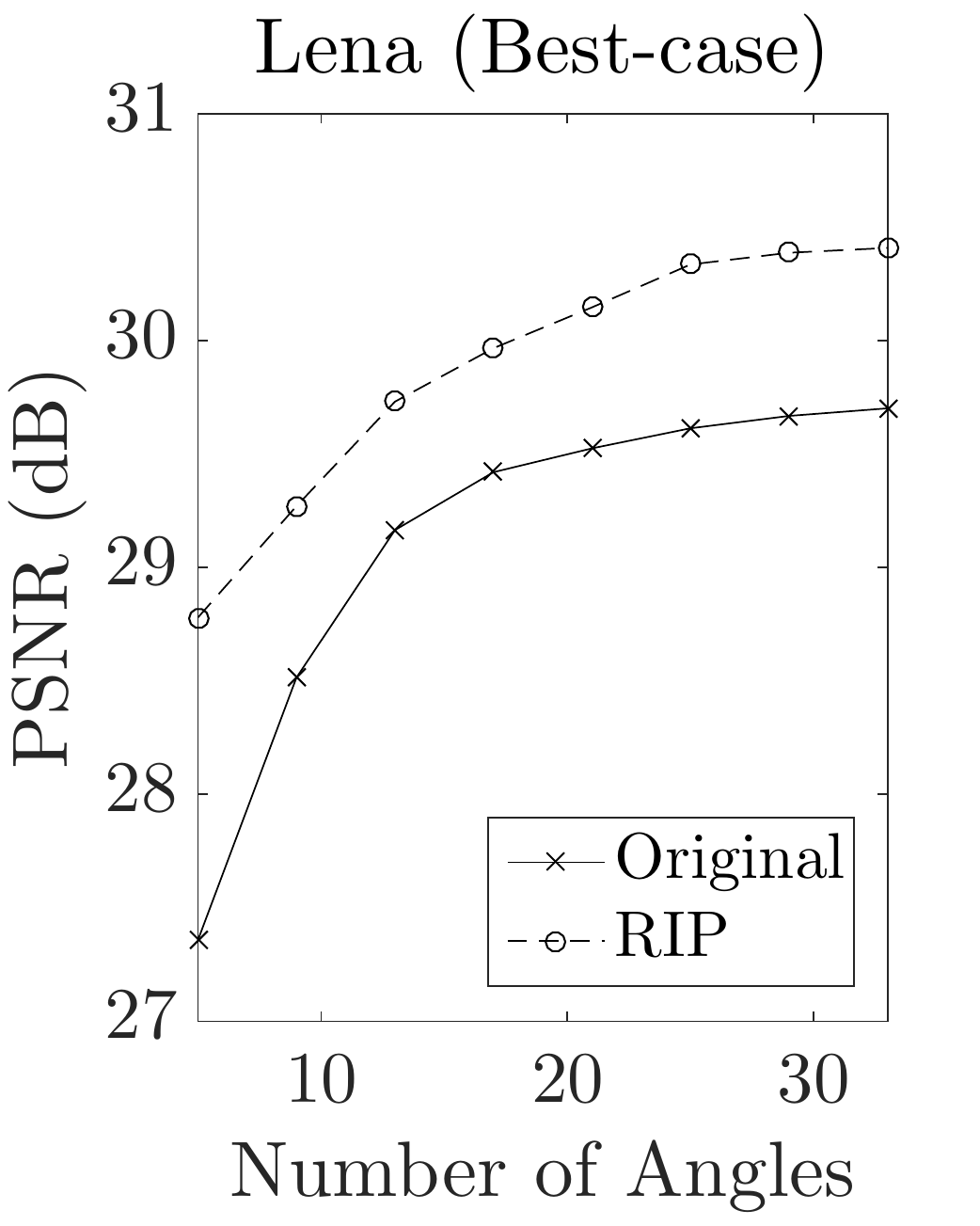}\includegraphics[width=0.333\linewidth]{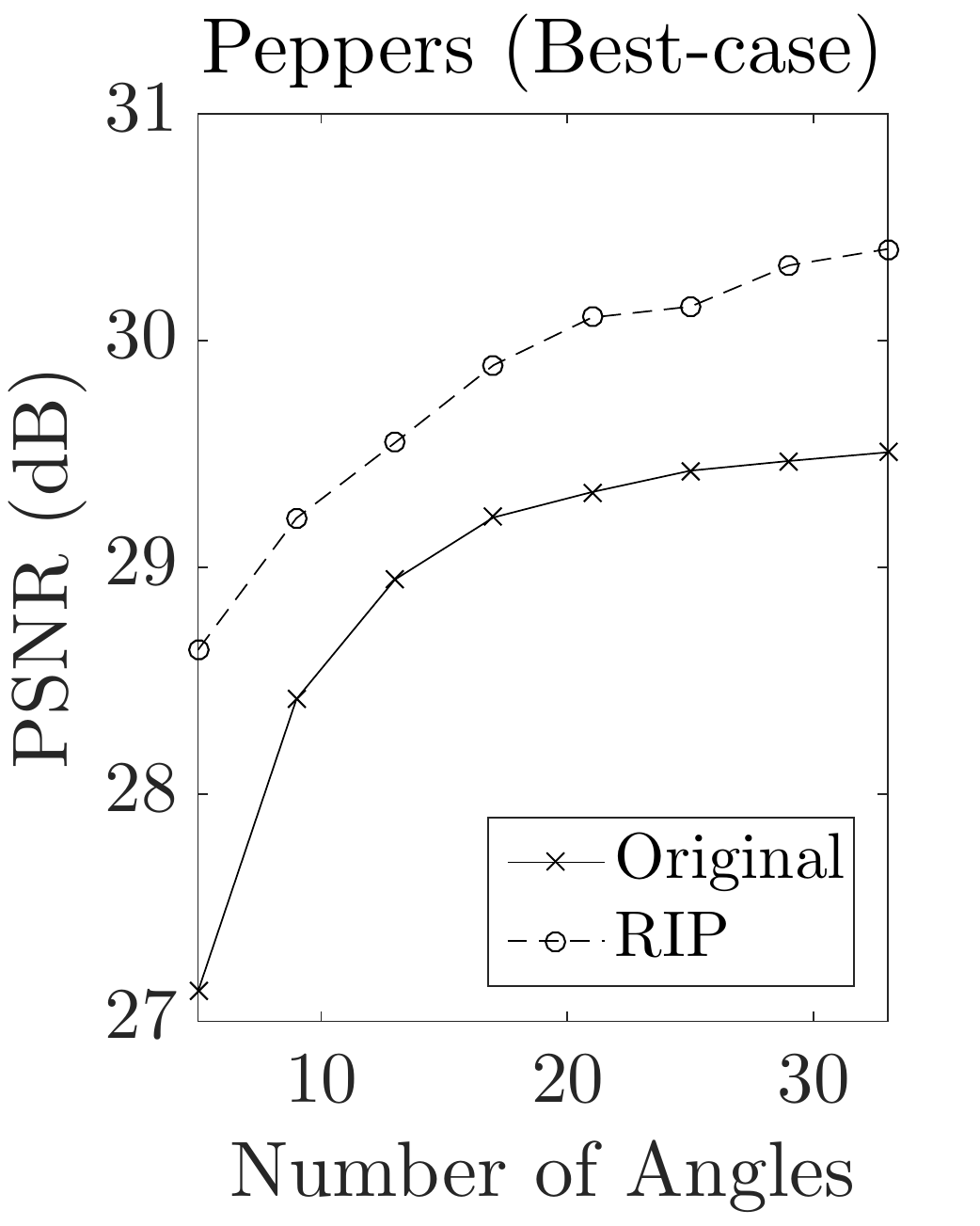}\includegraphics[width=0.333\linewidth]{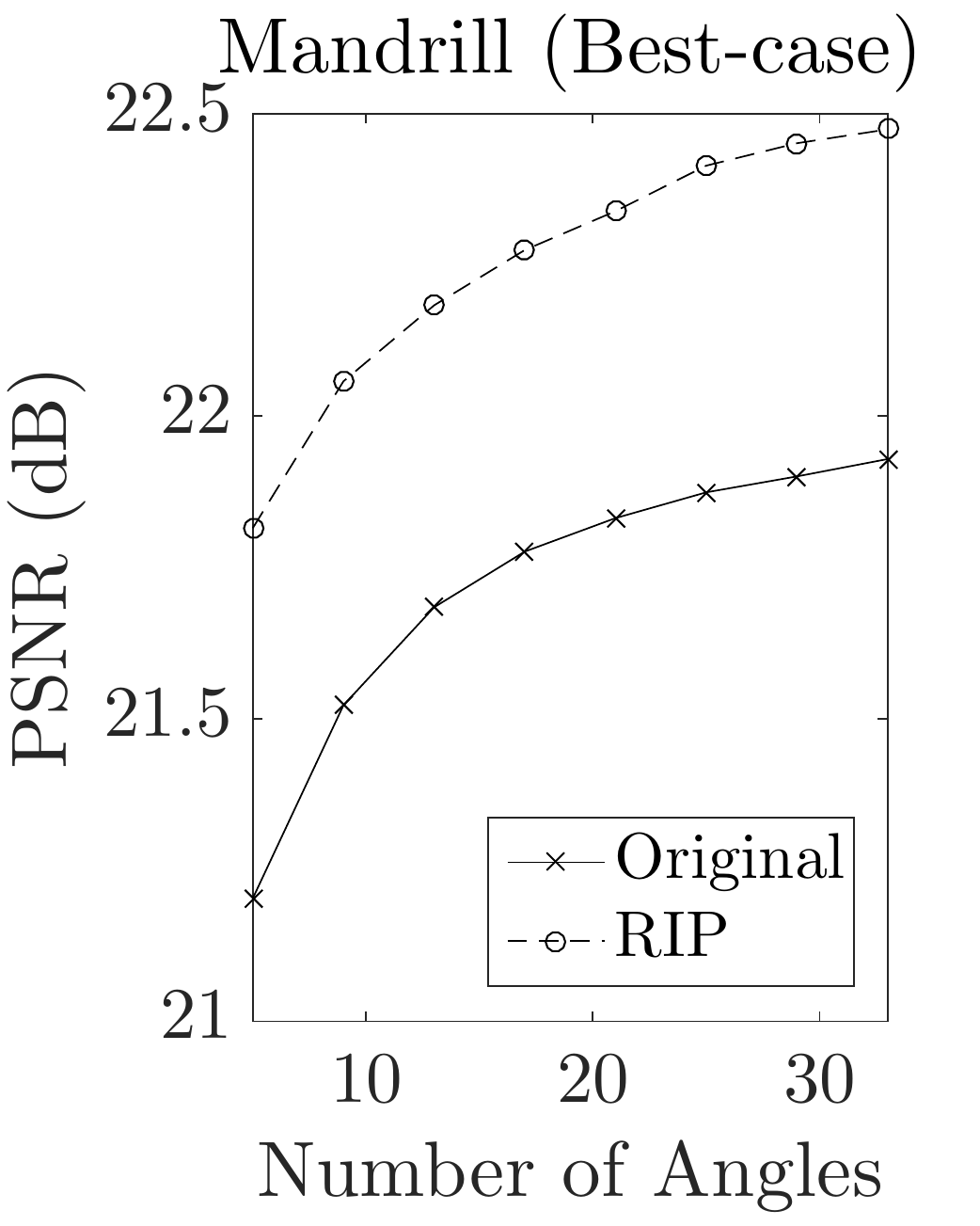}
	\caption{Best-case performance comparison between conventional and RIP predictions for various numbers of angular modes.}
\end{figure}

\begin{figure}[t!]
	\centering
	\includegraphics[width=0.333\linewidth]{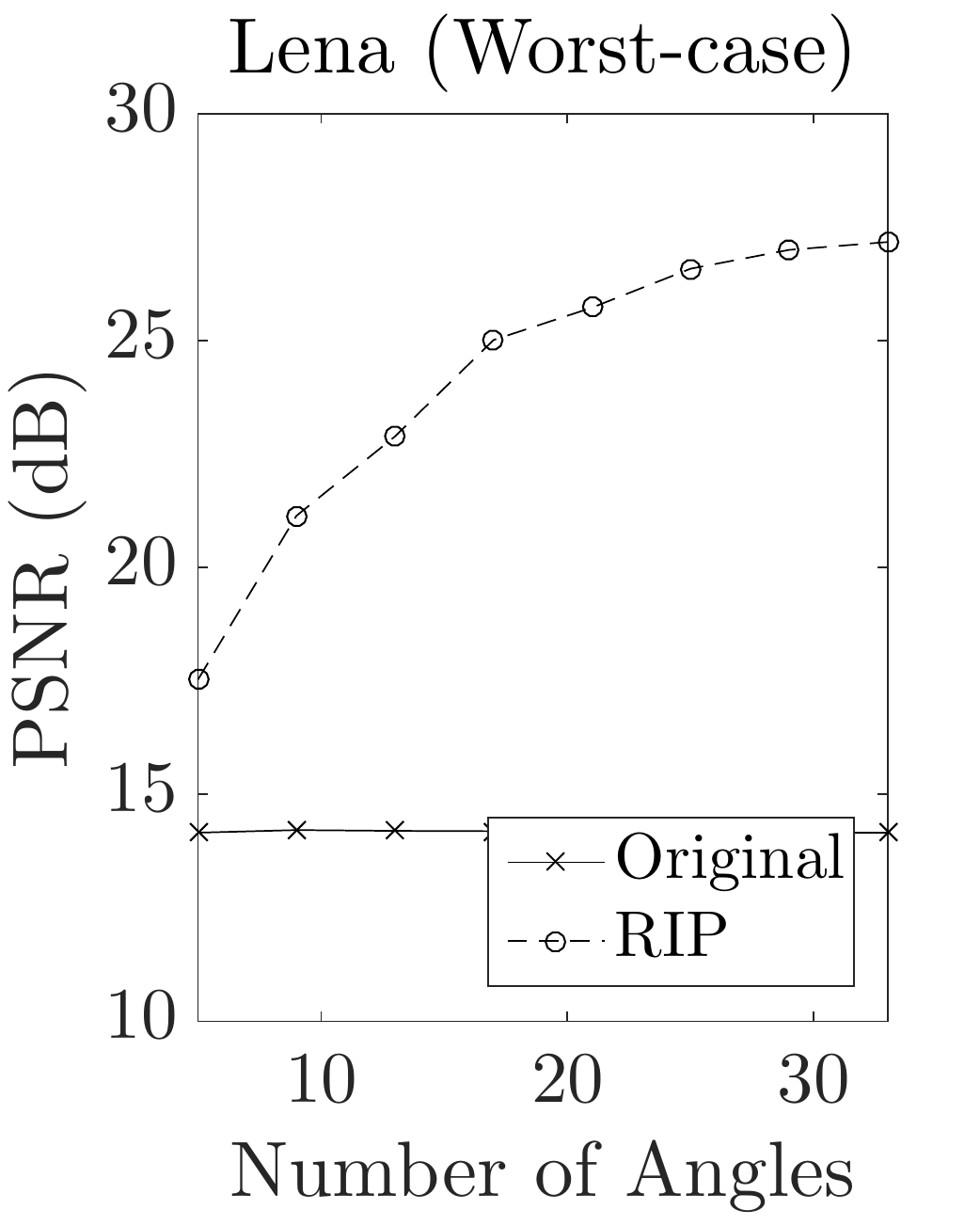}\includegraphics[width=0.333\linewidth]{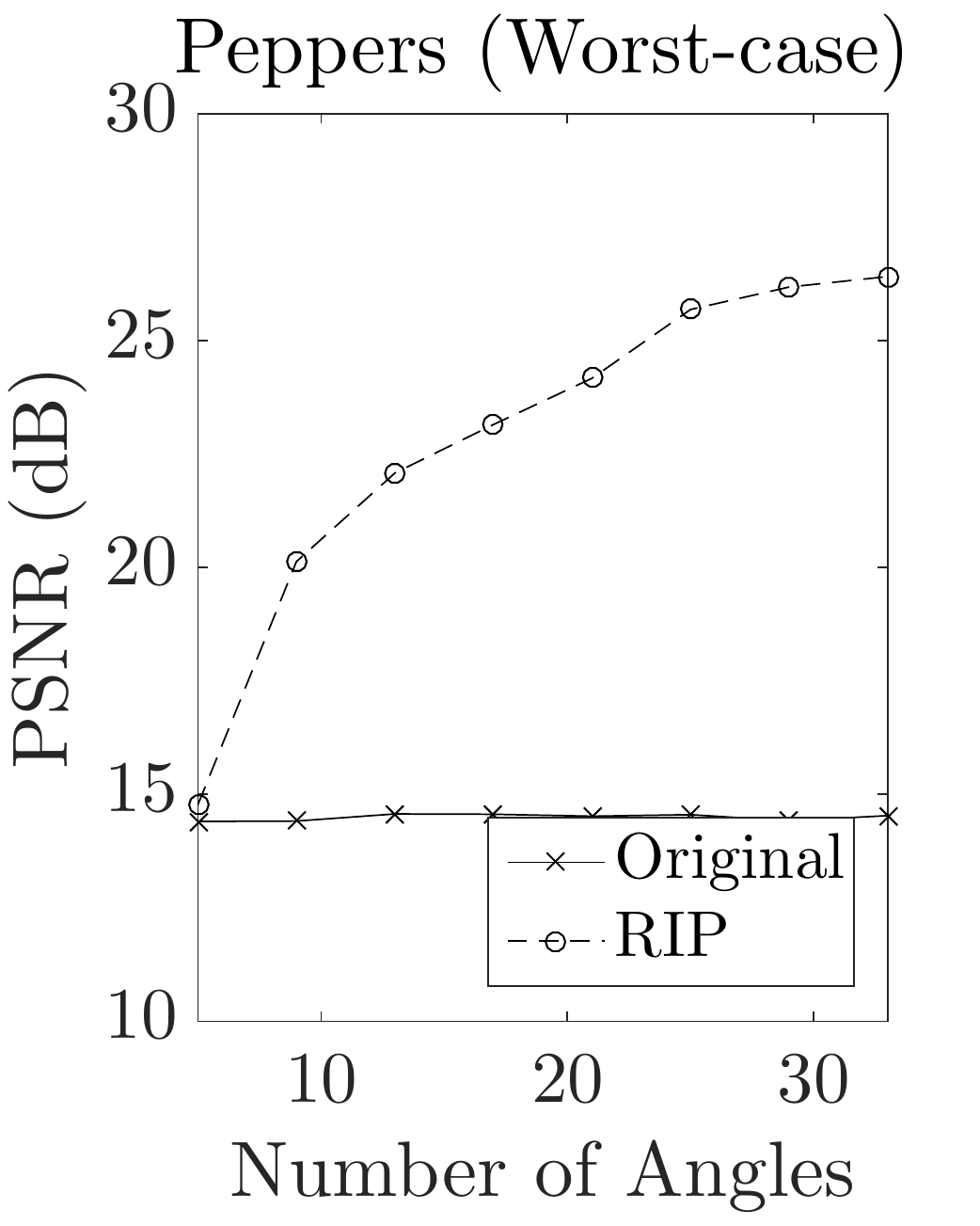}\includegraphics[width=0.333\linewidth]{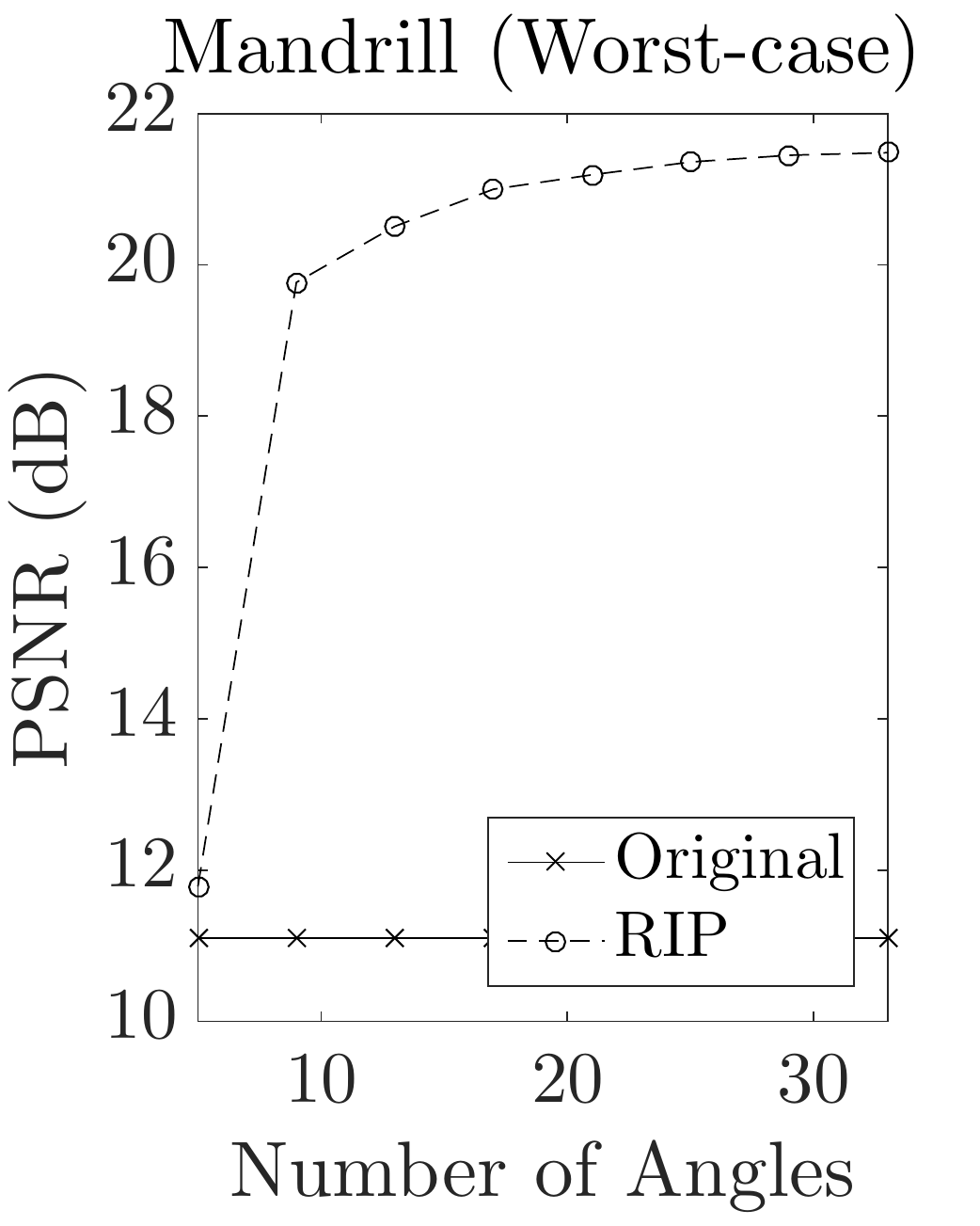}
	\caption{Worst-case performance comparison between conventional and RIP predictions for various numbers of angular modes.}
\end{figure}

The results for the best-case tests can be seen in Fig. 2. It is immediately apparent that the RIP estimates are superior across all test samples and in all modes. These results confirm our hypothesis that designed predictors, indeed, are not maximizing the natural image structure. Comparing the performance between the RIP and original estimates, we can observe a mean improvement of 0.757 dB, 0.854 dB, and 0.536 dB for the \emph{Lena}, \emph{Peppers}, and \emph{Mandrill} images respectively.

\begin{figure}[t!]
	\centering
	\frame{\includegraphics[width=0.5\linewidth]{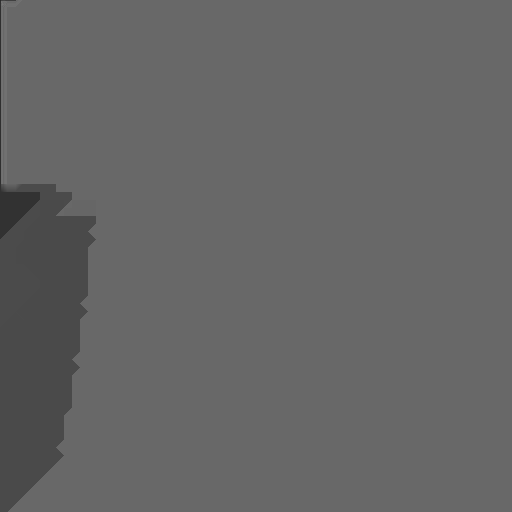}}\frame{\includegraphics[width=0.5\linewidth]{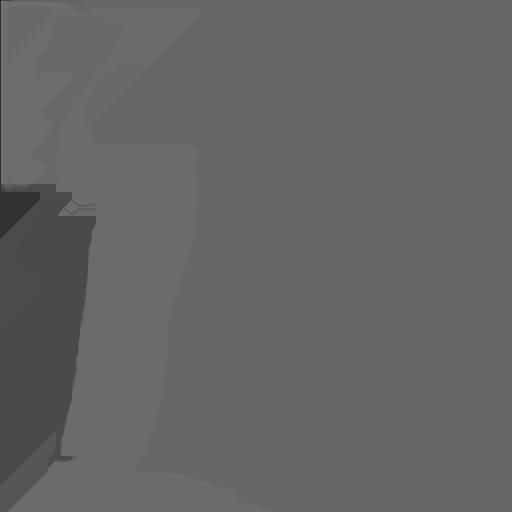}}
	\frame{\includegraphics[width=0.5\linewidth]{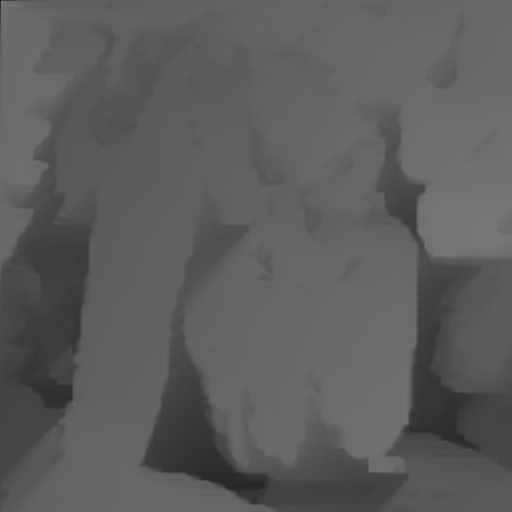}}\frame{\includegraphics[width=0.5\linewidth]{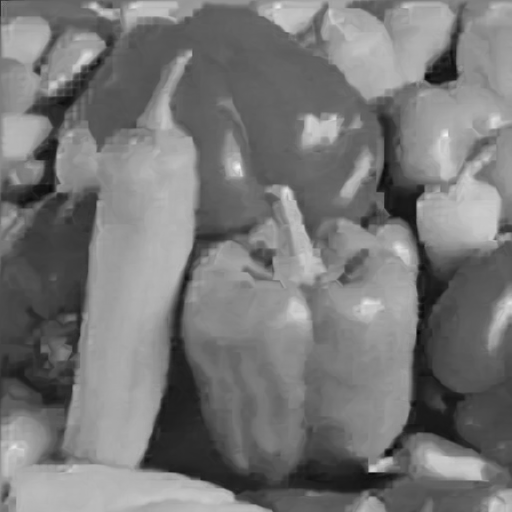}}
	\caption{Worst-case prediction with conventional (top row) and RIP (bottom row) schemes. RIP demonstrates a remarkable self-compensating property as it progress through the image. This behavior is less pronounced with 5 angular modes (left column) as compared with 33 angular modes (right column).}
\end{figure}

The worst-case performance results (seen in Fig. 3) demonstrate a surprising behavior. With conventional intra-prediction, pixels are copied and propagated throughout the image. With no error reporting, only pixels from the top-left patch can be propagated. On the other hand, with RIP, the use of regression matrices allows for fluctuations in the prediction values. These fluctuations allow for some form of prediction recovery even in the absence of additional data. This behavior can be more clearly seen in Fig. 4. With a low number of prediction modes, the system accomplishes a smooth but slow-reacting compensation. However, increasing the number of prediction modes allows for the selection of better models even in the most extreme cases of quantization. Compared to the best prediction, we observe an average loss of 5.747 dB, 6.693 dB, and 2.432 dB for the \emph{Lena}, \emph{Peppers}, and \emph{Mandrill} images respectively.

In addition to these results, a more complete visual comparison can be seen in Fig. 7 for the \emph{Peppers} image. This comparison illustrates the effect of varying the number of angular modes on the resulting best-case and worst-case estimates.

\subsection{Comparison with HEVC}
Our next series of experiments require the implementation of an HEVC-like prediction scheme. Here, we constructed a prediction matrix with all 33 angular modes as prescribed by the standard \cite{Lai2012}. The DC and planar prediction modes from HEVC are are also implemented. The resulting mappings are fed into our algorithm to produce a refined version. The best-case testing scheme described previously is used to evaluate and compare the predictor sets. We also provide data comparing prediction performance at $4 \times 4$, $8 \times 8$, $16 \times 16$, and $32 \times 32$ block sizes.

\begin{figure}[t!]
	\centering
	\includegraphics[width=0.333\linewidth]{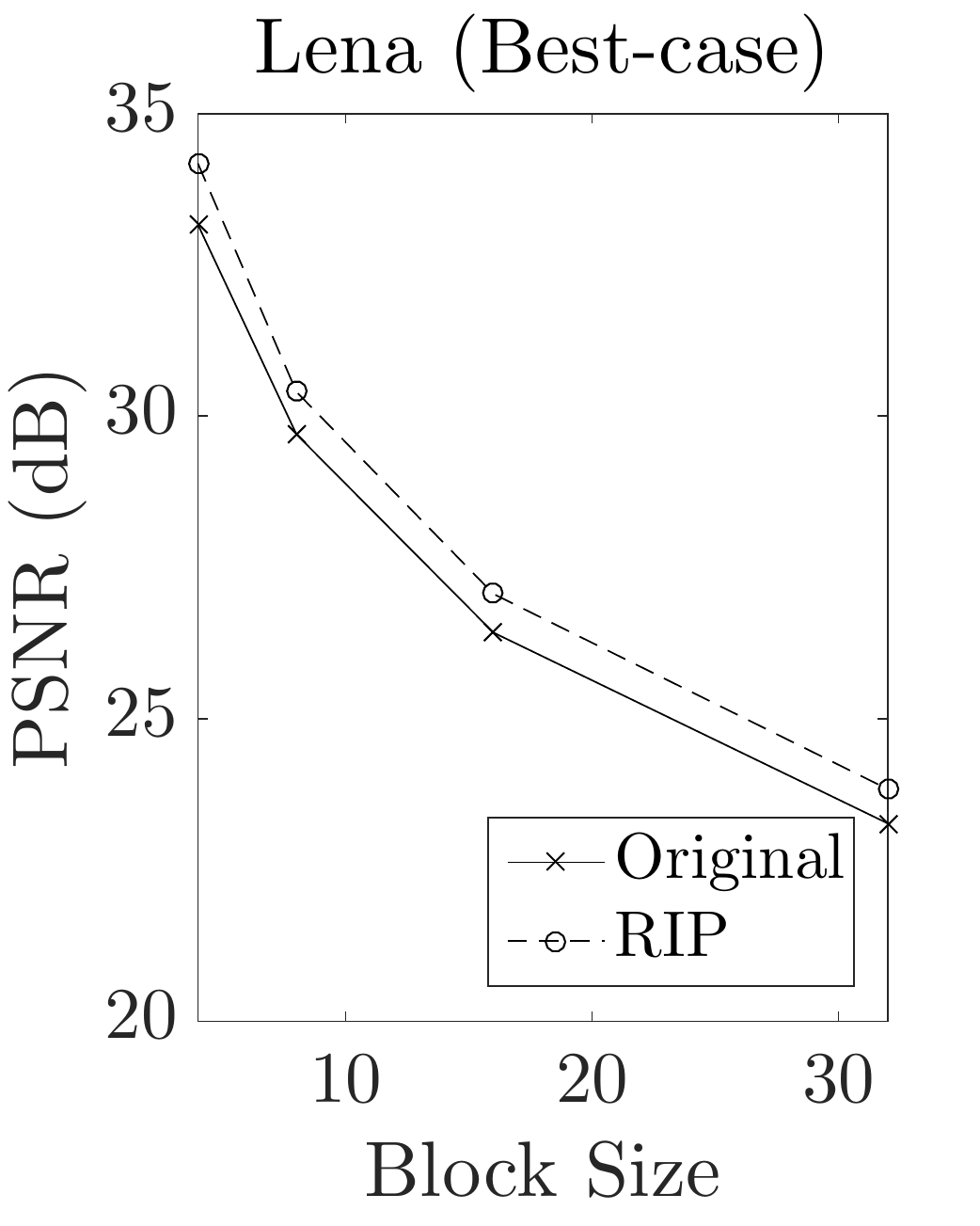}\includegraphics[width=0.333\linewidth]{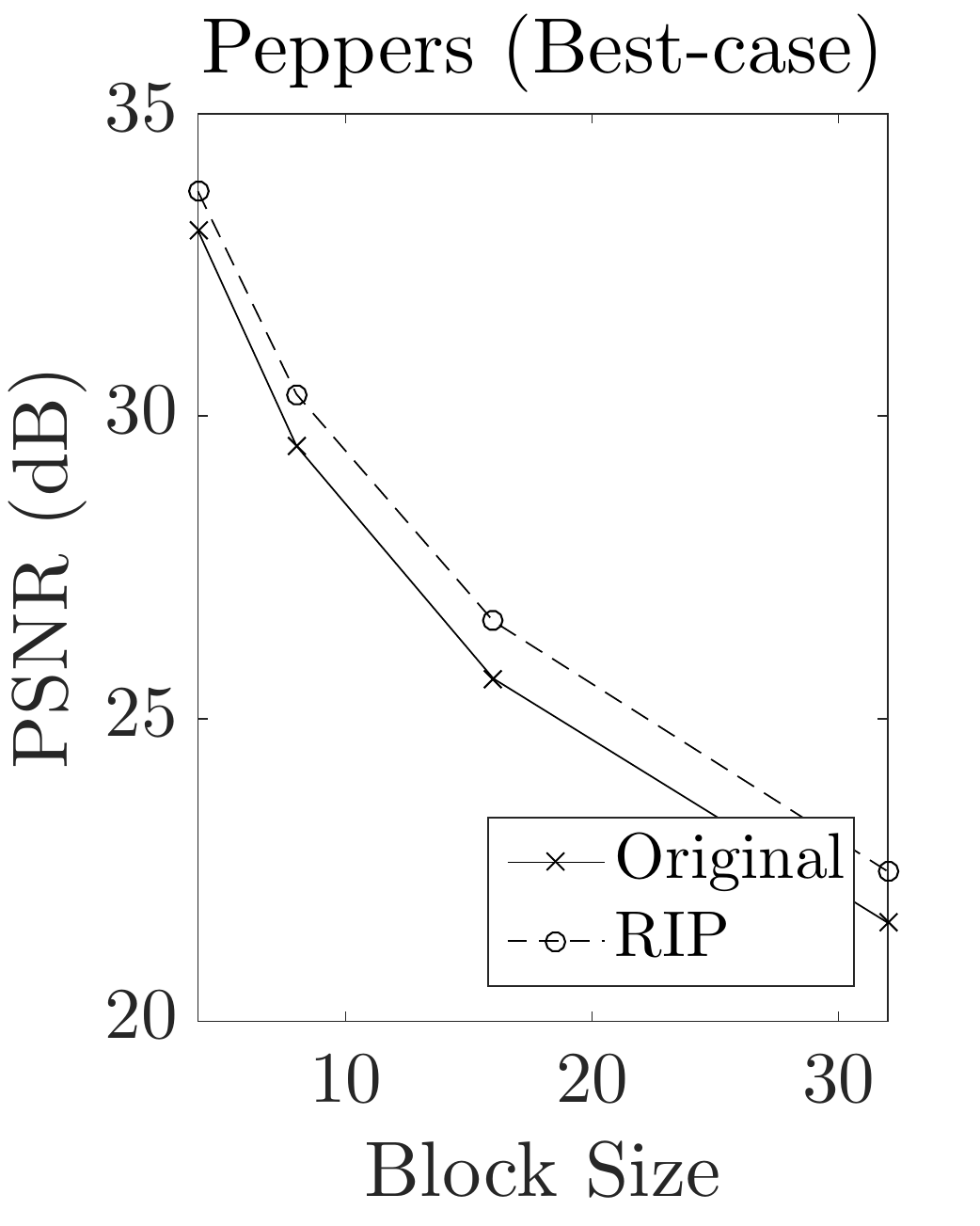}\includegraphics[width=0.333\linewidth]{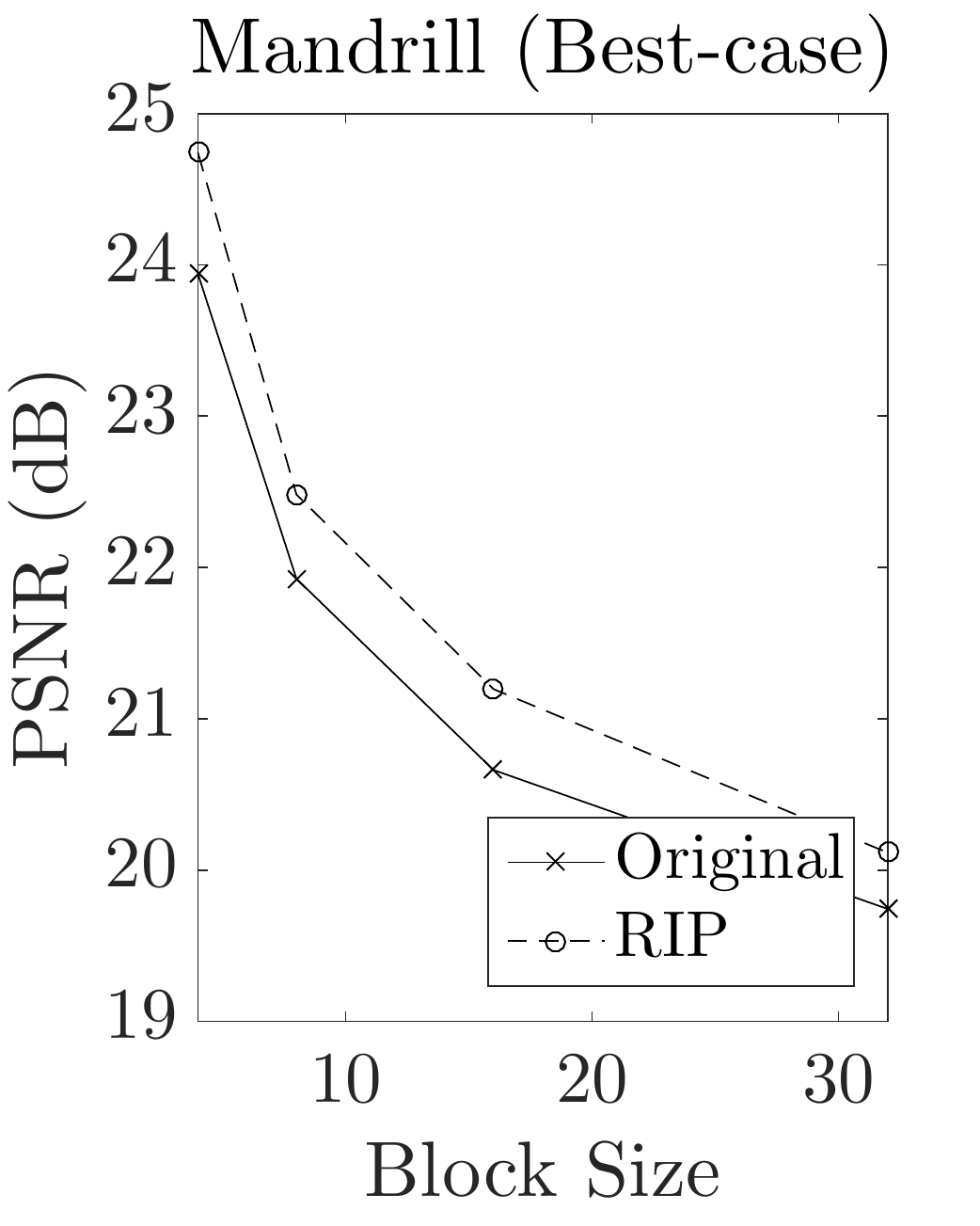}
	\caption{Best-case performance comparison between conventional and RIP predictions for various square block sizes.}
\end{figure}

\begin{figure}[t!]
	\centering
	\frame{\includegraphics[width=0.5\linewidth]{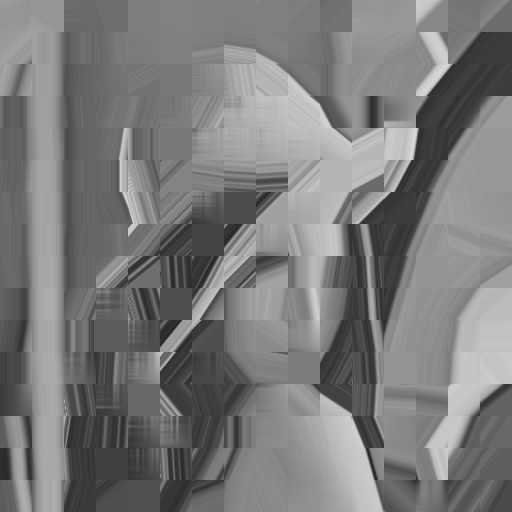}}\frame{\includegraphics[width=0.5\linewidth]{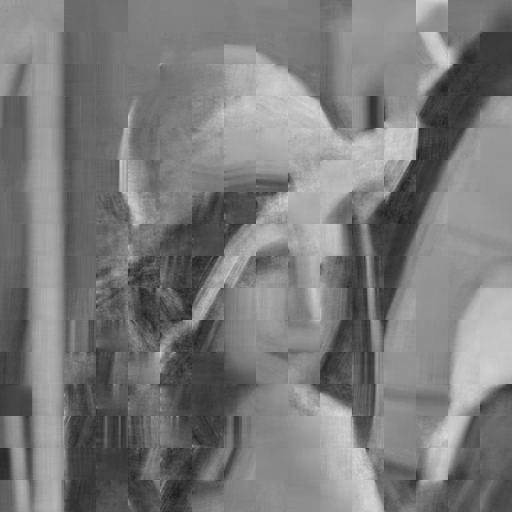}}
	\caption{Best-case prediction for a $32 \times 32$ block size on the \emph{Lena} image using conventional designed (left) and RIP (right) predictors.}
\end{figure}

The results from these particular tests (as seen in Fig. 5), once again demonstrate the general gains obtained when using RIP. As the block sizes increase, the RIP predictions maintain a strong quantitative lead over the HEVC-style predictions. However, from a subjective perspective (see Fig. 6), the resulting estimates may appear to be noisy. This is likely an effect of the increasing dimensionality gap between the reference inputs and the prediction pixels. For instance, in the $32 \times 32$ case, there are 1024 pixels to be predicted from 97 samples as compared to 64 predictions from 25 samples in the $8 \times 8$ case.

\subsection{Complexity}
An interesting note with regards to RIP is the complexity of the algorithm. Once all regressors have been trained, the intra-prediction task becomes a simple matrix multiplication operation. If we have $n$ pixels to predict from $m$ reference pixels, each group involves $mn$ multiplications. We can combine the prediction maps by stacking them as:
\begin{equation}
M = \left[
\begin{array}{c}
M_1\\
M_2\\
\vdots\\
M_k\\
\end{array}
\right]
\end{equation}
for all $k$ groups. Using this approach, all predictions can be made using a single matrix multiplication involving $kmn$ multiply-accumulate operations. While this approach is more computationally expensive than conventional intra-prediction, it has the advantage of being consistent. No special treatment is needed for any of the patches or any of the prediction modes. In addition, packages such as BLAS \cite{Kes2010} exploit optimizations to perform matrix multiplications \cite{Aga1994,Pot1996}. Modern hardware are often equipped to efficiently handle matrix-vector multiplications as well \cite{Kum2009,Kes2010}.

Another interesting observation in RIP is the capability to construct a good estimate of the image in the worst-case scenario. This would allow for a low-complexity coder that performs a serial sweep without error reporting. At this point, additional bits may then be transmitted describing the error between the worst-case reconstruction and the target image. Through such an approach, we completely bypass the RD decisions needed during the intra-coding step. The resulting errors can easily be divided into non-overlapping blocks and processed in parallel, leading to further computational reductions. While this approach is certainly sub-optimal, it nonetheless presents an interesting alternative to the traditional encoding scheme.

\begin{figure*}[t!]
	\centering
	\frame{\includegraphics[width=0.125\linewidth]{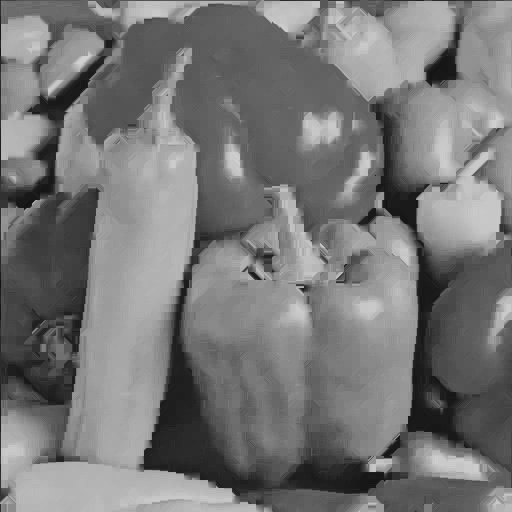}}\frame{\includegraphics[width=0.125\linewidth]{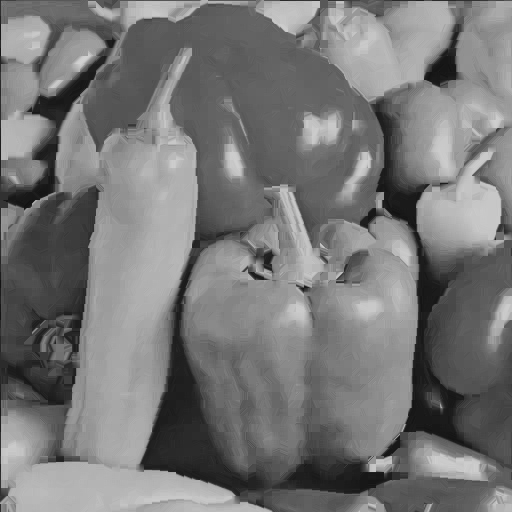}}\frame{\includegraphics[width=0.125\linewidth]{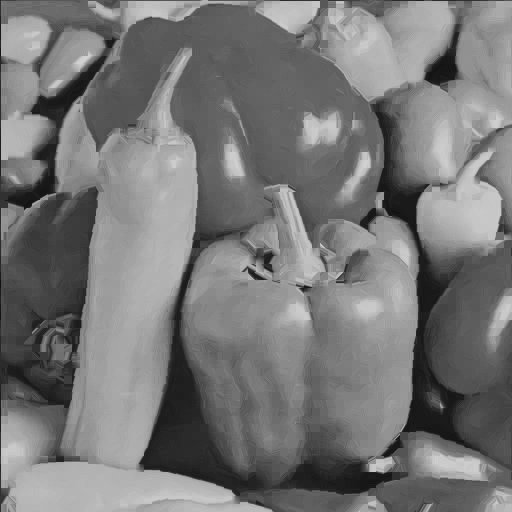}}\frame{\includegraphics[width=0.125\linewidth]{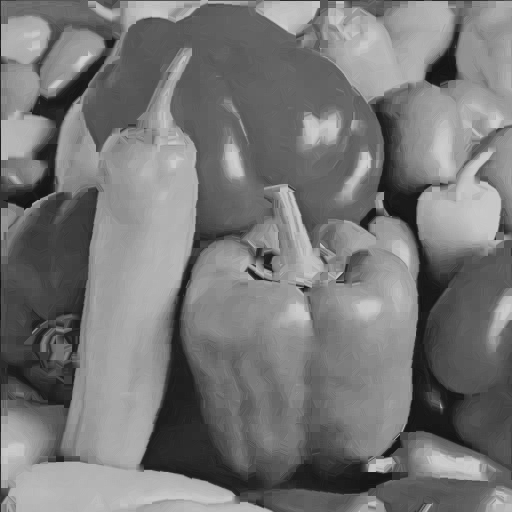}}\frame{\includegraphics[width=0.125\linewidth]{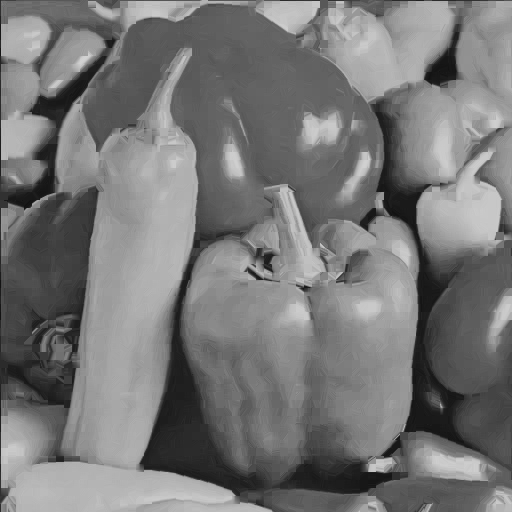}}\frame{\includegraphics[width=0.125\linewidth]{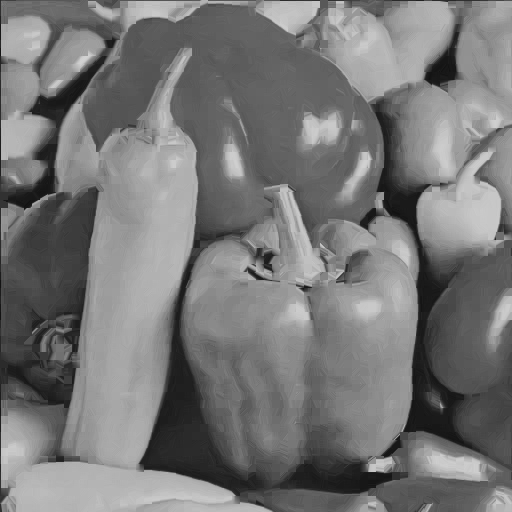}}\frame{\includegraphics[width=0.125\linewidth]{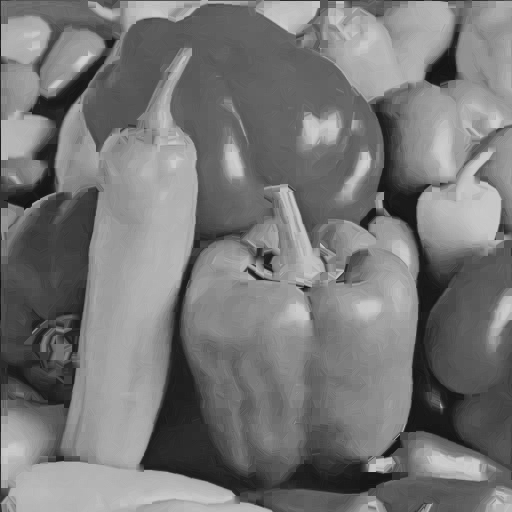}}\frame{\includegraphics[width=0.125\linewidth]{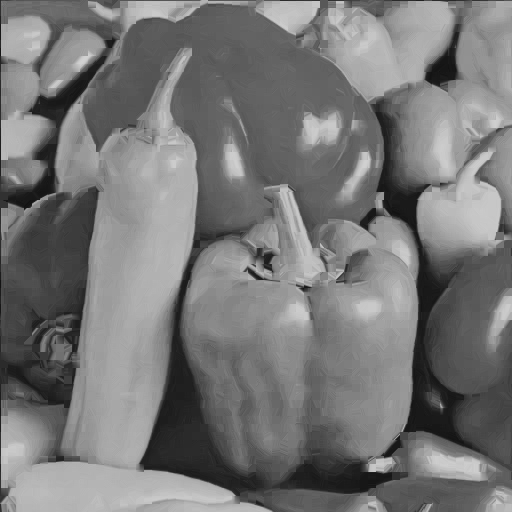}}
	\frame{\includegraphics[width=0.125\linewidth]{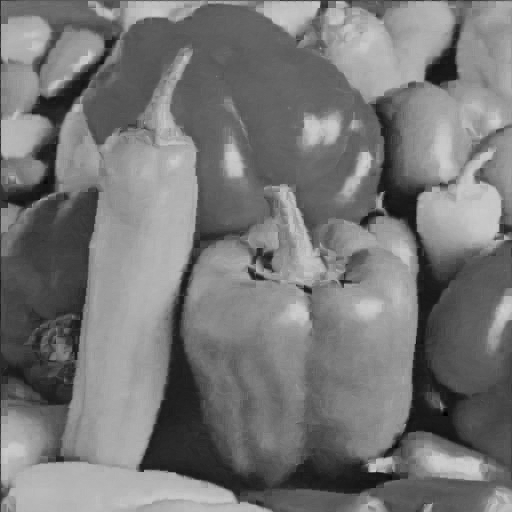}}\frame{\includegraphics[width=0.125\linewidth]{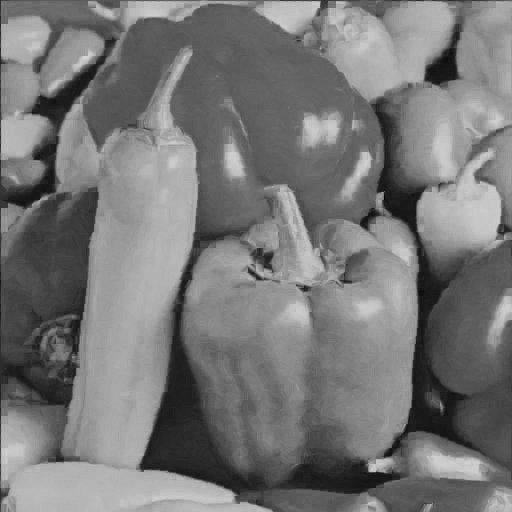}}\frame{\includegraphics[width=0.125\linewidth]{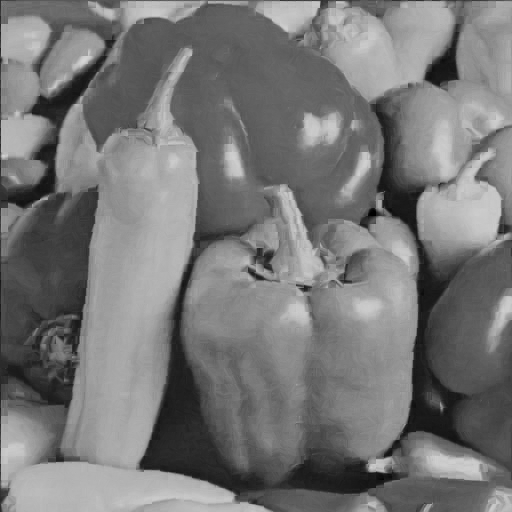}}\frame{\includegraphics[width=0.125\linewidth]{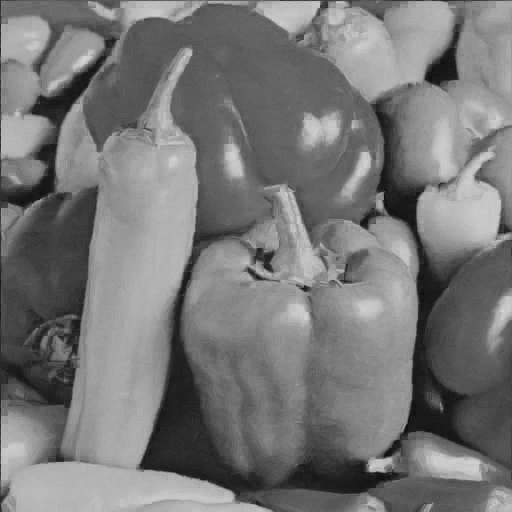}}\frame{\includegraphics[width=0.125\linewidth]{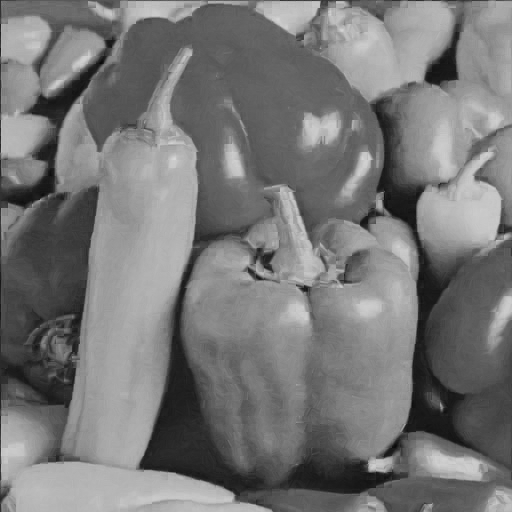}}\frame{\includegraphics[width=0.125\linewidth]{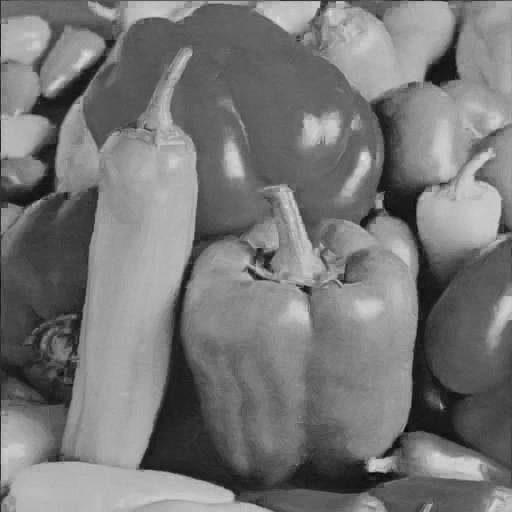}}\frame{\includegraphics[width=0.125\linewidth]{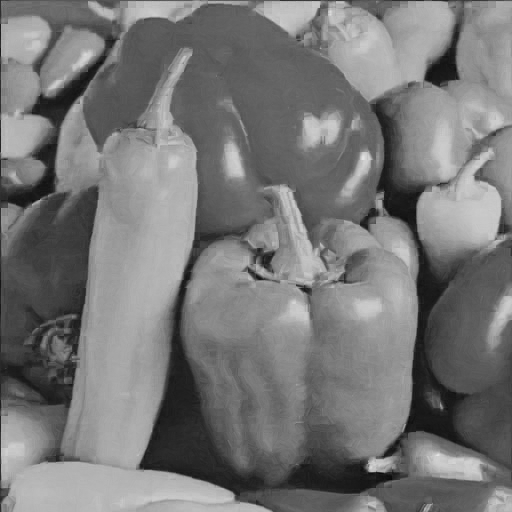}}\frame{\includegraphics[width=0.125\linewidth]{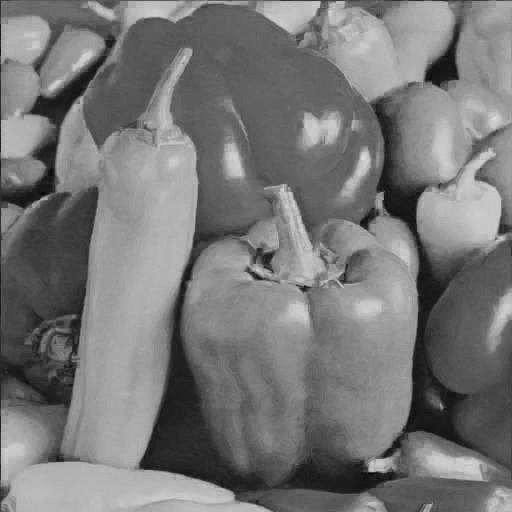}}
%	\frame{\includegraphics[width=0.125\linewidth]{peppers_5_angles_worst_orig.png}}\frame{\includegraphics[width=0.125\linewidth]{peppers_9_angles_worst_orig.png}}\frame{\includegraphics[width=0.125\linewidth]{peppers_13_angles_worst_orig.png}}\frame{\includegraphics[width=0.125\linewidth]{peppers_17_angles_worst_orig.png}}\frame{\includegraphics[width=0.125\linewidth]{peppers_21_angles_worst_orig.png}}\frame{\includegraphics[width=0.125\linewidth]{peppers_25_angles_worst_orig.png}}\frame{\includegraphics[width=0.125\linewidth]{peppers_29_angles_worst_orig.png}}\frame{\includegraphics[width=0.125\linewidth]{peppers_33_angles_worst_orig.png}}
	\frame{\includegraphics[width=0.125\linewidth]{peppers_5_angles_worst_ref.png}}\frame{\includegraphics[width=0.125\linewidth]{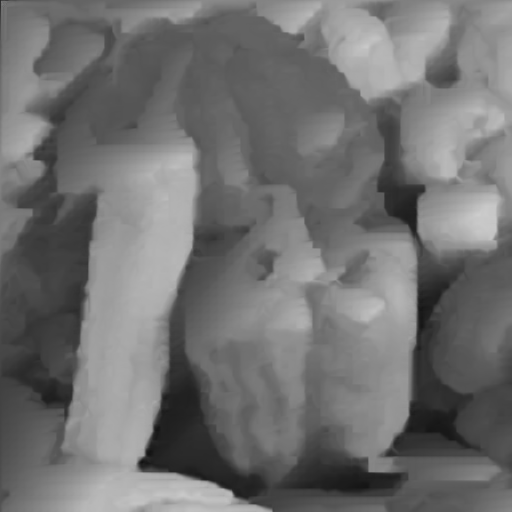}}\frame{\includegraphics[width=0.125\linewidth]{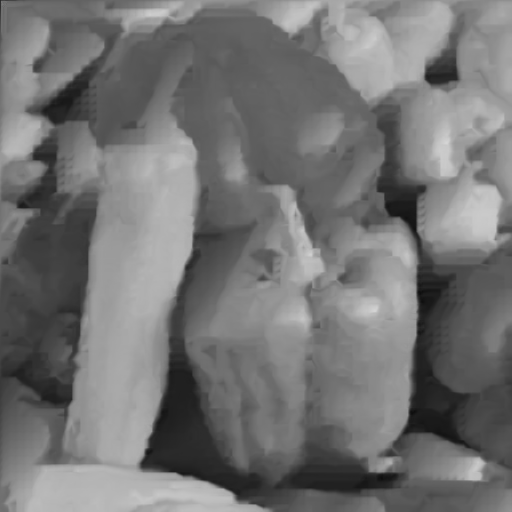}}\frame{\includegraphics[width=0.125\linewidth]{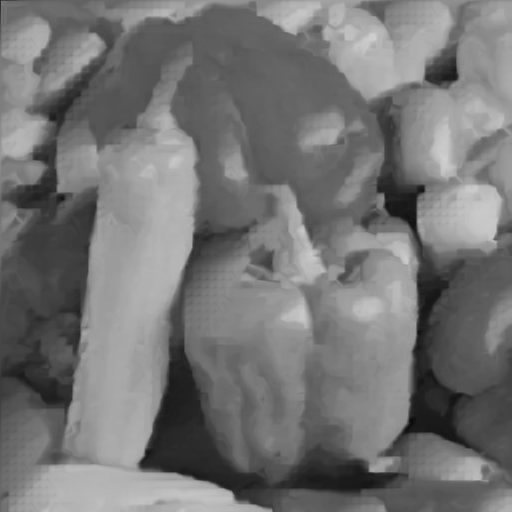}}\frame{\includegraphics[width=0.125\linewidth]{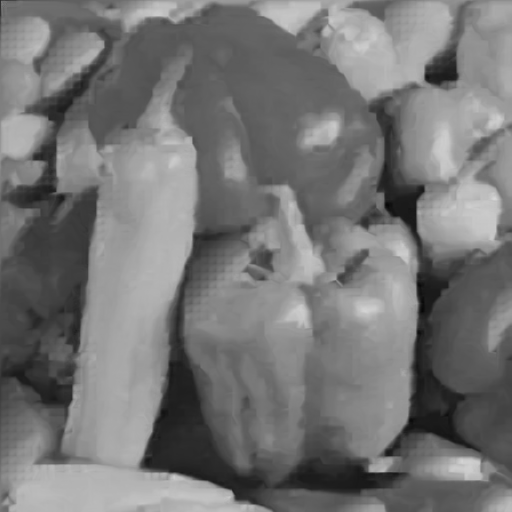}}\frame{\includegraphics[width=0.125\linewidth]{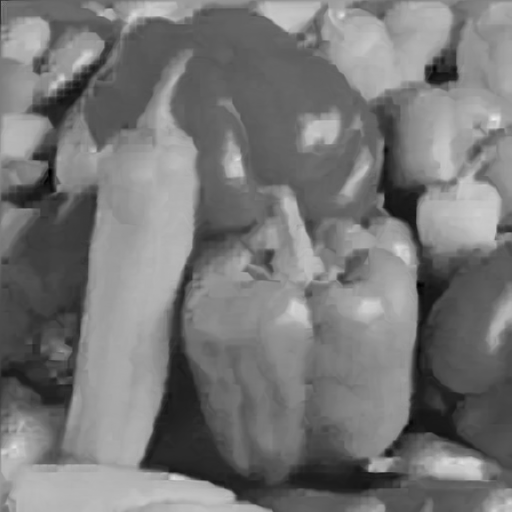}}\frame{\includegraphics[width=0.125\linewidth]{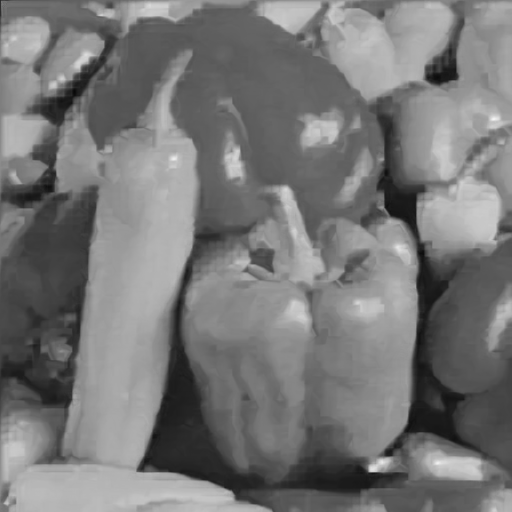}}\frame{\includegraphics[width=0.125\linewidth]{peppers_33_angles_best_ref.png}}
	\caption{Sample predictions using 5, 9, 13, 17, 21, 25, 29, and 33 angular modes (left to right) using an $8 \times 8$ block size. Illustrated from the top row to the bottom row respectively: best-case conventional prediction; best-case RIP prediction; worst-case RIP prediction.}
\end{figure*}

\section{Conclusion}
In this work, we formulated the intra-prediction task as a mapping problem between input reference pixels and the pixels to be predicted. In such a manner, we constructed an algorithm designed to refine an existing intra-prediction scheme. This refinement is performed iteratively using a regression-based approach leading to a significant improvement in the prediction quality compared to the HEVC prediction scheme. An interesting side effect of the RIP is that it is capable of compensating for the lack of information from previous pixels making it highly useful in low bitrate coding tasks.

% if have a single appendix:
%\appendix[Proof of the Zonklar Equations]
% or
%\appendix  % for no appendix heading
% do not use \section anymore after \appendix, only \section*
% is possibly needed

% use appendices with more than one appendix
% then use \section to start each appendix
% you must declare a \section before using any
% \subsection or using \label (\appendices by itself
% starts a section numbered zero.)
%

% Can use something like this to put references on a page
% by themselves when using endfloat and the captionsoff option.
\ifCLASSOPTIONcaptionsoff
  \newpage
\fi

% trigger a \newpage just before the given reference
% number - used to balance the columns on the last page
% adjust value as needed - may need to be readjusted if
% the document is modified later
%\IEEEtriggeratref{8}
% The "triggered" command can be changed if desired:
%\IEEEtriggercmd{\enlargethispage{-5in}}

% references section

% can use a bibliography generated by BibTeX as a .bbl file
% BibTeX documentation can be easily obtained at:
% http://mirror.ctan.org/biblio/bibtex/contrib/doc/
% The IEEEtran BibTeX style support page is at:
% http://www.michaelshell.org/tex/ieeetran/bibtex/
%\bibliographystyle{IEEEtran}
% argument is your BibTeX string definitions and bibliography database(s)
%\bibliography{IEEEabrv,TCSVT-Letters-arXiv}
%
% <OR> manually copy in the resultant .bbl file
% set second argument of \begin to the number of references
% (used to reserve space for the reference number labels box)
%\printbibliography
% biography section
% 
% If you have an EPS/PDF photo (graphicx package needed) extra braces are
% needed around the contents of the optional argument to biography to prevent
% the LaTeX parser from getting confused when it sees the complicated
% \includegraphics command within an optional argument. (You could create
% your own custom macro containing the \includegraphics command to make things
% simpler here.)
%\begin{IEEEbiography}[{\includegraphics[width=1in,height=1.25in,clip,keepaspectratio]{mshell}}]{Michael Shell}
% or if you just want to reserve a space for a photo:

\begin{IEEEbiography}{Carlo Noel Ochotorena}
received the B.S. and M.S. degrees in Electronics and Communications Engineering from De La Salle University, Manila, Philippines in 2009 and 2012 respectively.

Since 2013, he has been a professor with the Electronics and Communications Engineering at De La Salle University, Manila Philippines. He is also currently a Doctoral student at the Tokyo Institute of Technology. His research interests include image processing, compressive sensing, and computational photography.
\end{IEEEbiography}

\begin{IEEEbiography}{Yukihiko Yamashita}
was born in 1960 in Kanagawa Prefecture, Japan. He received the B.E., the M.E., and the Dr. Eng. degrees from Tokyo Institute of Technology in 1983, 1985, and 1993, respectively.

From 1985 to 1988 he was with the Japan Atomic Energy Research Institute. From 1988 to 1989 he was with the ISAC corporation. In 1989 he joined the faculty of the Tokyo Institute of Technology, where he is now an associate professor of the Course of Engineering.

His research interests include pattern recognition and image processing.

He received a Paper Award in 1993 from the Institute of Electronics, Information, and Communication Engineers of Japan (IEICE).

Dr. Yamashita is a member of the Institute of Electrical and Electronics Engineers, the Institute of Electronics, Information, and Communication Engineers of Japan, and Information Processing Society of Japan.
\end{IEEEbiography}

\vfill
% if you will not have a photo at all:
%\begin{IEEEbiographynophoto}{John Doe}
%Biography text here.
%\end{IEEEbiographynophoto}

% insert where needed to balance the two columns on the last page with
% biographies
%\newpage

%\begin{IEEEbiographynophoto}{Jane Doe}
%Biography text here.
%\end{IEEEbiographynophoto}

% You can push biographies down or up by placing
% a \vfill before or after them. The appropriate
% use of \vfill depends on what kind of text is
% on the last page and whether or not the columns
% are being equalized.

%\vfill

% Can be used to pull up biographies so that the bottom of the last one
% is flush with the other column.
%\enlargethispage{-5in}

% that's all folks
\end{document}